\documentclass[%
reprint,
superscriptaddress,
nofootinbib,
 amsmath,amssymb,
 aps,
]{revtex4-2}

\usepackage{graphicx}
\usepackage{dcolumn}
\usepackage{bm}
\usepackage{braket}
\usepackage{upgreek}
\usepackage{xcolor}
\usepackage{afterpage}
\usepackage{relsize}
\usepackage{multirow}
\usepackage{float}
\usepackage{physics}
\usepackage{makecell}
\usepackage{placeins}
\usepackage[colorlinks=true,urlcolor=blueprl,citecolor=blueprl,linkcolor=blueprl]{hyperref}
\usepackage{lineno}
\usepackage[symbol]{footmisc}
\usepackage{hhline}

\newcommand{\ozlem}[1]{\textcolor{black}{#1}}
\newcommand{\ozlemblue}[1]{\textcolor{black}{#1}}
\newcommand{\ozlemREV}[1]{\textcolor{black}{#1}}

\newcolumntype{P}[1]{>{\centering\arraybackslash}p{#1}}

\usepackage{hyperref}
\let\svthefootnote\thefootnote
\newcommand\freefootnote[1]{%
  \let\thefootnote\relax%
  \footnotetext{#1}%
  \let\thefootnote\svthefootnote%
}

\definecolor{blueprl}{RGB}{46,48,146}

\def\UQA{Centre for Quantum Computation and Communication Technology, School of Mathematics and Physics, University of Queensland, St Lucia, QLD 4072, Australia}
\def\ASTAR{A*STAR Quantum Innovation Centre (Q.InC), Agency~for~Science,~Technology~and~Research~(A*STAR), 2 Fusionopolis Way, Innovis \#08-03, Singapore 138634, Republic of Singapore}

\begin{document}


\title{Software-enhanced simultaneous quantum-classical communication protocol with Gaussian post-selection}
\author{\"{O}zlem Erk{\i}l{\i}\c{c}$^{*}$}
\freefootnote{$^*$ \href{ozlemerkilic1995@gmail.com}{ozlemerkilic1995@gmail.com}}
\affiliation{\UQA}

\author{Biveen Shajilal}
\affiliation{\ASTAR}

\author{Nicholas Zaunders}
\affiliation{\UQA}

\author{Timothy C. Ralph}
\affiliation{\UQA}

\date{\today}
\begin{abstract}
    Simultaneous quantum-classical communication~(SQCC) protocols offer a practical approach to continuous-variable quantum key distribution~(CV-QKD) by encoding quantum and classical signals onto the same optical pulse. However, like most QKD protocols, their performance is limited when experimental parameters, such as modulation variance, are optimised based on stationary channel assumptions. In fluctuating environments, such as free-space links, this can result in sub-optimal key rates and reduced transmission distances. In this work, we \ozlemblue{introduce Gaussian post-selection into the SQCC framework}, enabling a software-based optimisation of the modulation variance after channel estimation. This passive approach enhances key rates in both asymptotic and finite-size regimes without requiring hardware modifications \ozlem{and remains effective even when receiver imperfections are taken into account.} We demonstrate that our protocol improves the transmission distance and robustness of SQCC \ozlemREV{relative to the standard fixed-variance SQCC protocol}, and approaches the performance of a fully pre-optimised system across both fibre and free-space channels. In particular, we show that the protocol enables full communication windows under \ozlemblue{ideal weather conditions and maintains higher duty cycles during adverse weather} in satellite-to-ground scenarios. These results highlight the practicality of post-selection based SQCC for real-world quantum communication over both terrestrial fibre networks and satellite-based free-space links.
\end{abstract}
             
\maketitle

\section*{\label{sec:introduction}Introduction}
Quantum key distribution~(QKD) enables two remote parties to generate a shared secret key with security guaranteed by the principles of quantum mechanics~\cite{gisin2002,ekert2014,pirandola2020advances,usenko2025continuous}. One variant of QKD uses continuous variables~\cite{ralph1999,hillery2000}, leveraging the amplitude and phase quadratures of light to encode key information and offering a practical route to integration with conventional optical communication systems. The quadratures are then measured using homodyne or heterodyne detection. One \ozlemblue{popular scheme of} continuous-variable QKD (CV-QKD) involves Gaussian encoding, where coherent states are modulated according to a Gaussian distribution~\cite{grosshans2002continuous,grosshans2003quantum,weedbrook2004quantum,lance2005no,lodewyck2007quantum,madsen2012continuous,jouguet2013experimental,wang201525,jain2022practical,hajomer2024long}. When the modulation variance is optimised for the channel characteristics to obtain the highest key rates possible, this protocol is commonly referred to as GG02~\cite{grosshans2002continuous,grosshans2003quantum,weedbrook2004quantum}. In most QKD implementations, a parameter estimation is carried out to quantify channel loss and noise, which in turn bounds Eve’s information~\cite{garcia2007quantum} and used to tune the experimental parameters. While effective under stable conditions, this process assumes that the estimated parameters remain valid throughout \ozlemblue{the transmission of the key}. However, in highly dynamic environments, such as free-space channels, the fluctuations can cause these \ozlemblue{estimated} parameters to be obsolete and 
\ozlemblue{compromise the security, effectively reducing the key rates.}

\ozlemblue{While demonstrations of QKD over large networks have confirmed its scalability~\cite{peev2009secoqc, chen2010metropolitan, sasaki2011field, pirandola2013high, chen2021integrated, chen2021implementation, hajomer2025high}, practical deployment in real-world communication systems requires the coexistence of quantum and classical signals within the same infrastructure. A natural way to perform this is to multiplex classical and quantum signals by leveraging wavelength division multiplexing~\cite{chapuran2009optical, eraerds2010quantum, patel2014quantum} and polarisation multiplexing techniques~\cite{khaksar2023simultaneous}. These hybrid approaches enable cost-effective integration, seamless upgrade paths, and the immediate extension of quantum communication capabilities to metropolitan and transnational networks. As a result, quantum-enhanced security can be deployed over practical infrastructure, accelerating the real-world adoption and impact of QKD.}

\ozlemblue{An even more hardware-efficient approach for integrating CV-QKD into classical communication channels was proposed by Qi~\textit{et al.}~\cite{qi2016simultaneous} \ozlemREV{and experimantally tested~\cite{kumar2019experimental}}, referred to as simultaneous quantum-classical communication~(SQCC). In this scheme,} quantum information is first encoded by preparing coherent states modulated from a bivariate Gaussian distribution. Classical information is then embedded by applying a large displacement, chosen from a discrete classical alphabet, to each quantum state~\cite{qi2016simultaneous, zaunders2024quantum, zaunders2025enhanced}. After the channel, Bob performs a heterodyne detection and applies a redisplacement to extract the quantum information. This approach takes a step toward real-world deployment by eliminating the need for separate channels for classical and quantum communication, enabling both to coexist on the same optical link. While this scheme offers practical advantages, existing SQCC protocols~\cite{qi2016simultaneous,qi2018noise,pan2020simultaneous,zaunders2024quantum,winnel2024classical} typically adopt simplified modelling assumptions and do not comprehensively account for finite-size effects. A recent protocol by Zaunders~\textit{et al.}~\cite{zaunders2025enhanced} introduced an improved SQCC model with composable finite-size security, showing that classical bit errors \ozlem{leads to a non-physical state between Alice and Bob, which was corrected} via a renormalisation step. However, the fundamental challenges of CV-QKD persist in SQCC protocols. Each set of channel parameters requires an optimal modulation variance, and even with channel characterisation, these estimates can quickly become unreliable in varying conditions.

Various passive techniques have been explored in CV-QKD to increase key rates without relying on active feedback, including quantum processes such as noiseless linear amplification~\cite{xiang2010heralded, winnel2020generalized}, phase-sensitive amplification~\cite{bencheikh2001quantum}, photon subtraction~\cite{huang2013performance, hu2020continuous}, photon addition~\cite{chen2021continuous}, and photon catalysis~\cite{ye2019improvement, hu2020continuous, hu2021performance}. Although these methods can enhance the performance, their experimental realisation is challenging. An alternative to these physical processes involves a method called post-selection, where applying a quantum process is equivalent to first detecting the quantum state followed by post-processing of the measured outcomes with a digital filter. For instance, measurement-based NLA (MB-NLA) imitates the effect of a physical NLA by applying Gaussian post-selection to the measurement results. This technique has been shown to extend the transmission range of CV-QKD~\cite{walk2013security, fiuravsek2012gaussian, hosseinidehaj2020finite} and improve performance in CV quantum communication protocols~\cite{chrzanowski2014measurement, zhao2017characterization, zhao2023enhancing, shajilal2024improving}. This approach has also been used for photon subtraction, photon addition and photon catalysis by using appropriate filters to emulate. For instance, photon subtraction and photon addition processes are replaced with a non-Gaussian virtual photon subtraction~\cite{li2016non, zhong2018self} and photon addition~\cite{jeng2025entanglement} filters, respectively. Similarly, the photon catalysis is imitated by a Gaussian zero-photon catalysis post-selection~\cite{zhong2020virtual}. However, since all of these post-selection filters imitate physical operations, their parameters are still confined by these processes. 
\begin{figure*}[t!]
\center{\includegraphics[scale=0.32]
        {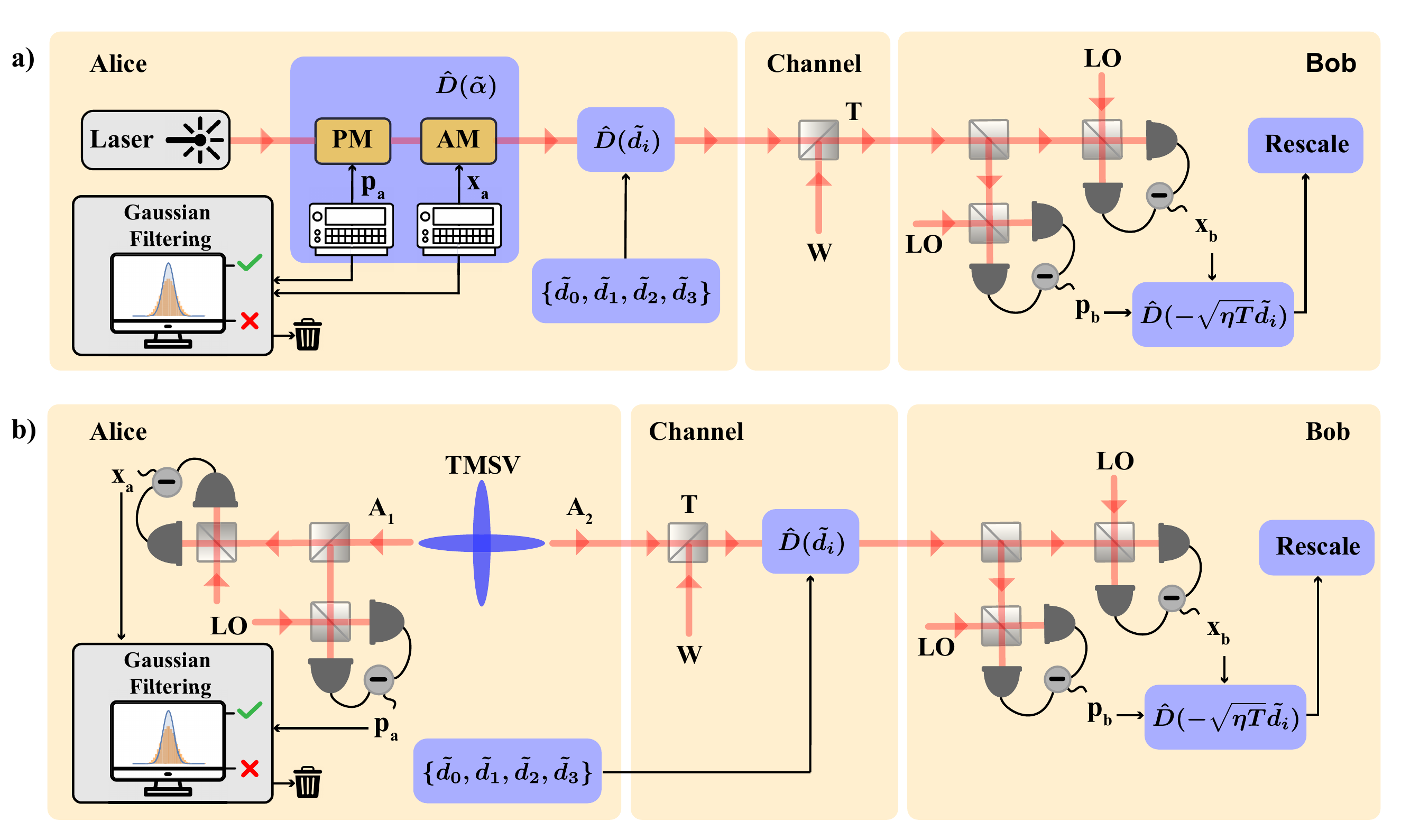}}
\caption{\label{fig:protocol_figure}Schematic of the filtering process in the SQCC protocol. \textbf{(a)} Prepare-and-measure model where Alice encodes coherent states $\ket{(x_a+ip_a)/2}=\ket{\Tilde{\alpha}}$ from a Gausssian distribution with zero mean and variance, $V_{mod}$, using white noise generated by function generators. She samples $x_a$ and $p_a$ simultaneously, then applies a large phase displacement of $\Tilde{d}$ chosen from her alphabet $d_i$. The encoded states are then sent through a quantum channel with transmittivity $T$ and thermal noise $W$. Bob performs heterodyne detection to measure $x_b$ and $p_b$, simultaneously and re-displaces his measurement outcomes by $\Tilde{d}$. If deemed necessary, he applies an electronic gain to rescale his data to make the joint distribution between Alice and Bob Gaussian. After data acquisition and once the states have been sent to Bob, Alice applies a Gaussian filter to her quantum data kept on her computer. \textbf{(b)} Entanglement-based model where Alice prepares a two-mode squeezed vacuum (TMSV) state, retains mode $A_1$ and performs a heterodyne measurement. Mode $A_2$ is sent to Bob through a quantum channel. A classical displacement operation is applied after the channel, under the assumption that Eve has full knowledge of the classical communication. Bob performs a heterodyne measurement on the received mode, re-displaces the outcomes, and applies a renormalisation procedure to restore a Gaussian distribution. Alice and Bob then estimate the channel parameters. If Alice’s modulation variance is found to be sub-optimal for the estimated channel, she applies a Gaussian filter to her heterodyne outcomes $x_a$ and $p_a$, effectively emulating an optimally prepared state. PM/AM: Electro-optic phase/amplitude modulators. LO: Local oscillator.
}
\end{figure*}

Recently Erk{\i}l{\i}\c{c} \textit{et al.}~\cite{erkilic2025enhanced} proposed a post-selection method for CV-QKD with Gaussian modulation and homodyne or heterodyne detection, which avoids the need to emulate any underlying physical operation for filtering. The Gaussian filter used in this paper is effectively equivalent to modifying the modulation variance of the protocol via post-selection to match the variance of the optimal GG02 protocol~\cite{grosshans2002continuous,grosshans2003quantum,weedbrook2004quantum}. This is particularly useful for fluctuating channels, where the modulation variance cannot be pre-optimised as the channel conditions may have changed since the last characterisation, potentially resulting in sub-optimal or even zero key rates. \ozlemblue{This approach, however, did not address finite-size composable security, which is essential for assessing realistic implementations.}

In this work, \ozlem{we integrate the Gaussian filtering approach of Erk{\i}l{\i}\c{c} \textit{et al.}\cite{erkilic2025enhanced} into the improved SQCC protocol of Zaunders~\textit{et al.}~\cite{zaunders2025enhanced}, creating a unified framework for post-selected SQCC.} \ozlemblue{In contrast to the CV-QKD case~\cite{erkilic2025enhanced}, the inclusion of post-selection within SQCC requires a distinct security analysis that accounts for additional error sources and the renormalisation steps inherent to the SQCC protocol.} \ozlem{We further extend the analysis to more realistic scenarios by incorporating detector inefficiency and electronic noise at the receiver,} \ozlemblue{effects that were not included in the SQCC model of Zaunders~\textit{et al.}~\cite{zaunders2025enhanced}.} The filtering is performed after Bob rescales his data and applies a renormalisation step to restore \ozlem{physical} statistics. Alice then applies post-selection to her modulation data and communicates the retained indices to Bob through a standard sifting procedure. This Gaussian-filtered SQCC protocol offers a practical and passive optimisation method for SQCC and enables modulation variance tuning after transmission for fluctuating channels such as satellite and free-space links. This work also incorporates a finite-size composable security analysis tailored to the Gaussian post-selection applied within the SQCC framework, \ozlemblue{extending beyond the asymptotic treatment in the earlier CV-QKD study~\cite{erkilic2025enhanced}.}

\section{\label{sec:protocol}Protocol}
The main protocol builds on the SQCC scheme introduced by Zaunders \textit{et al.}~\cite{zaunders2025enhanced}, with the key addition of a Gaussian post-selection applied to Alice’s measurement outcomes, as illustrated in Fig.~\ref{fig:protocol_figure}. In QKD, the prepare-and-measure (PM) scheme is commonly used, where Alice prepares a coherent state and sends it to Bob, who performs a heterodyne measurement. This approach is equivalent to the entanglement-based (EB) scheme, in which Alice generates a two-mode squeezed vacuum (TMSV) state, keeps one mode for heterodyne detection, and sends the other to Bob~\cite{grosshans2003virtual, garcia2007quantum, weedbrook2012gaussian}. The protocol in the prepare-and-measure scheme can be summarised as:
\begin{enumerate}
    \item Alice encodes coherent states $\ket{(x_a+ip_a)/2}=\ket{\Tilde{\alpha}}$ from a bivariate Gaussian distribution with zero mean and variance, $V_{mod}$, corresponding to the quantum symbols for the key generation. She keeps a copy of her encoding to perform post-selection after estimating the channel parameters if deemed necessary.
    \item She then displaces her states by some large phase displacement, $\Tilde{d}$, randomly chosen from her classical alphabet ${d_i}$.
    \item The state $\ket{\Tilde{\alpha}+\Tilde{d}}$ representing both the classical and quantum symbols, is then sent through a thermal-loss channel modelled with transmittivity $T$ and thermal-noise $\xi$.
    \item Bob performs a heterodyne detection on the received signal to obtain $\Tilde{\beta}$ which is a Gaussian variable with a mean value of $\sqrt{\eta T}d$ where $\eta$ represents his detection efficiency.
    \item Bob then re-displaces his measurement outcomes by $\Tilde{d}$ linked to the classical information to retrieve the quantum information. This maps the protocol to an equivalent GG02 protocol with coherent states encoding and heterodyne measurements.
    \item As discussed in Zaunders \textit{et al.}~\cite{zaunders2025enhanced}, any mistakes in the re-displacement operation of Bob result in a non-physical distribution between Alice and Bob. In the case of such mistake, Bob can rescale his data by applying an electronic gain, denoted as $N_d$, to obtain an equivalent \ozlem{physical non-Gaussian} distribution.
    \item Alice and Bob then estimate the channel parameters. If the modulation variance of the quantum data is sub-optimal, causing Eve’s inferred information to exceed their mutual information, Alice applies a Gaussian filter to her data, following the method of Erk{\i}l{\i}\c{c} \textit{et al.}~\cite{erkilic2025enhanced} to optimise the modulation and enable key generation.
\end{enumerate}

\section{\label{sec:analysis}Security Analysis}
The SQCC protocol was first introduced by Qi~\textit{et al.}~\cite{qi2016simultaneous}, with several other protocols developed since then that follow the same core principle of simultaneous quantum and classical communication~\cite{qi2018noise, pan2020simultaneous, zaunders2024quantum, winnel2024classical}. However, these early models all assumed that any errors Bob makes during the classical re-displacement would manifest as added Gaussian noise. In reality, such errors introduce non-physical features into the measurement distribution. This issue was later addressed by Zaunders~\textit{et al.}~\cite{zaunders2025enhanced}, who proposed rescaling Bob’s measurement outcomes using an electronic gain to \ozlem{restore physicality, leading to a non-Gaussian distribution in the end.} The initial EB covariance matrix between Alice and Bob after Bob rescales his data can be expressed as
\begin{equation}
    \label{eq:initial_CM_EB}
    \sigma_{AB}=\begin{pmatrix}
    V\mathbb{I}_2 & N_dC_d\sigma_z\\
    N_dC_d\sigma_z & V_b\mathbb{I}_2
    \end{pmatrix},
\end{equation}
\ozlemblue{where $\sigma_z=\mathrm{diag}(1,-1)$.} $V=V_{mod}+1$, with $V_{mod}$ is Alice's initial modulation variance. $C_d$, and $V_{b}$ represent Alice and Bob's covariance and Bob's initial variance before rescaling respectively. $N_d$ denotes the electronic gain applied by Bob to his measurement outcomes in order to recover a Gaussian distribution. These parameters are given as
\begin{equation}
\label{eq:covariance_AB}
    C_d=\sqrt{\eta T(V^2-1)}(1-\delta),
\end{equation}
\begin{equation}
\label{eq:bobs_variance}
    V_b=\eta\big(TV+(1-T)W\big)+(1-\eta)+2v_{el},
\end{equation}
\begin{equation}
\label{eq:electronic_gain}
    N_d=\sqrt{\frac{V_{b}+1}{V_{b_d}+1}}, 
\end{equation}
\begin{equation}
\label{eq:bobs_variance_rescaled}
    V_{b_d}=V_b+2\alpha^2e_C-2(V_b+1)\delta-2\alpha^2e_C^2,
\end{equation}
 where $V_{b_d}$ represents Bob's variance after rescaling. $T$ and $W$ denote the channel transmission and thermal noise, respectively, with $W=\xi T/(1-T)+1$ for an excess noise of $\xi$. $\eta$ and $v_{el}$ are Bob's detection efficiency and detector noise, respectively. $\alpha=\sqrt{\eta T}d$ is the re-displacement that Bob applies after Alice's signal has gone through the channel, while $d$ is Alice's large displacement. $e_C$ represents the bit-error rate of the classical signal, where $e_C$ and $\delta$ are given by
\begin{equation}
\label{eq:bit_error_rate}
    e_C=\frac{1}{2}\mathrm{erfc}\bigg(\frac{\sqrt{\mathrm{SNR}}}{2}\bigg),
\end{equation}
\begin{equation}
    \label{eq:delta_eqn}
    \delta=\sqrt{\frac{\mathrm{SNR}}{\pi}}e^{-\mathrm{SNR}/4},
\end{equation}
where $\mathrm{SNR}$ denotes the signal-to-noise ratio as
\begin{equation}
    \label{eq:SNR}
    \mathrm{SNR}=\frac{\alpha^2}{V_b+1}.
\end{equation}

To further optimise the protocol, a Gaussian post-selection is applied on Alice's side using the filter,
\begin{equation}
    \label{eq:alices_filter}
    F_A(x_a,p_a)=e^{-g^2(x_a^2+p_a^2)},
\end{equation}
where  $x_a$ and $p_a$ are the randomly generated variable and  \text{$g\geq0$} represents the filter gain. \ozlemREV{This filter is designed such that for all $g \ge 0$, the resulting state shared between Alice and Bob remains physical: when $g = 0$ the covariance matrix is unchanged, while for $g > 0$ the modulation variance decreases but never below the vacuum limit.} Each symbol, $x_a$ and $p_a$, are either kept or discarded based on the probability $F_A(x_a,p_a)$. Note that in the EB scheme, Alice applies the same filter with an equivalent gain, $g'$. The random variables generated by Alice's function generator follow a Gaussian distribution as described below
\begin{equation}
    \label{eq:alices_dist}
    P_a(x_a,p_a)=\frac{1}{2\pi V_{mod}}\mathrm{exp}\bigg[\frac{-(x_a^2+p_a^2)}{2V_{mod}}\bigg],
\end{equation}
where $V_{mod}$ represents the modulation variance of Alice. When Alice applies the filter shown in Eq.~\eqref{eq:alices_filter}, the probability of success is given by

\begin{equation}
    \label{eq:alices_ps}
    P_{A}\!=\!\!\int_{-\infty}^\infty \!\int_{-\infty}^\infty\!\!\!\!\!F_A(x_a,p_a) P_a(x_a,p_a)\,d{x_a}d{p_a}\!=\!\frac{1}{2g^2V_{mod}+1}.
\end{equation}
After Alice's post-selection, the modulation variance of the effective thermal state is reduced. The equivalent modulation variance is expressed as follows
\begin{align}
    \label{eq:new_modulation_variance}
    \Tilde{V}_{mod}&=\int_{-\infty}^\infty\int_{-\infty}^\infty\!\!\frac{x_a^2F_A(x_a,p_a)P_a(x_a,p_a)}{P_{A}}\,d{p_a}\,d{x_a}\nonumber\\
    &=\frac{V_{mod}}{2g^2 V_{mod}+1}.
\end{align}
Alice's post-selection cannot modify the channel parameters. However, Bob's effective variance also reduces, as he must disregard the data that Alice discards. Therefore, Bob's variance, using Alice's new variance $\Tilde{V}=\Tilde{V}_{mod}+1$, can be expressed as
\begin{equation}
    \label{eq:bobs_variance2}
    \Tilde{V}_{b}=\eta\big(T\Tilde{V}+(1-T)W\big)+(1-\eta)+2v_{el}.
\end{equation}
Similarly, Alice and Bob's covariance also changes,
\begin{equation}
    \label{eq:new_covariance}
    \Tilde{C}_d=\sqrt{\eta T(\Tilde{V}^2-1)}(1-\delta).
\end{equation}
Therefore, the post-selected covariance matrix between Alice and Bob becomes
\begin{equation}
    \label{eq:postselected_matrix}
    \Tilde{\sigma}_{AB}=\begin{pmatrix}
    \Tilde{V}\mathbb{I}_2 & N_{d}\Tilde{C}_d\sigma_z\\
    N_{d}\Tilde{C}_d\sigma_z & \Tilde{V}_b\mathbb{I}_2
    \end{pmatrix}=\begin{pmatrix}
        \Tilde{a}\mathbb{I}_2 & \Tilde{c}\sigma_z \\
        \Tilde{c}\sigma_z & \Tilde{b}\mathbb{I}_2
    \end{pmatrix}.
\end{equation}
Note that the scaling factor $N_d$ uses the initial variances of Bob as this quantity is a physical rescaling that Bob has to do in order to make their data Gaussian. This process is done prior to the post-selection.

\subsection{Asymptotic Key Rate}
The asymptotic key rate is computed using
\begin{equation}
    K^\infty=P_A(\beta I_{AB}-I_E),
\end{equation}
where $I_{AB}$ represents Alice and Bob's mutual information while $I_E$ is Eve's Holevo bound. $\beta$ is the reconciliation efficiency, with $0 \leq \beta < 1$. As Bob performs heterodyne detection, Alice and Bob's mutual information is obtained from
\begin{equation}
    \label{eq:mutual information}
    I_{AB}=\log_2\bigg(\frac{V_A}{V_{A|B}}\bigg),
\end{equation}
where $V_A=(\Tilde{a}+1)/2$ is Alice's variance of her heterodyne measurement in the equivalent EB scheme. $V_{A|B}$ denotes Alice's variance conditioned on Bob's measurement outcomes and calculated from
\begin{equation}
    \label{eq:conditional_variance}
    V_{A|B}=V_A-\frac{\phi^2}{V_B},
\end{equation}
where $V_B=(\Tilde{b}+1)/2$ is the variance of Bob's measurement outcomes and Alice and Bob's covariance after their measurements scales as $\phi=\Tilde{c}/2$.

Eve's information is bounded using the Holevo quantity~\cite{holevo1998capacity}, which provides an upper bound on the amount of information she can access. Since the quantum channel is Gaussian and a Gaussian filter is applied on Alice's side, we use the Gaussian extremality theorem, which states that Gaussian states maximise the von Neumann entropy for a fixed covariance matrix~\cite{garcia2006unconditional, wolf2006extremality}. It is important to note that although Eve is assumed to have full knowledge of the classical displacement, she does not have access to Bob’s detection efficiency. Even if she were aware of its value, she would not be able to control it, as doing so would require access to Bob’s laboratory, an assumption incompatible with the basic security premise of QKD. Similarly, Eve is not assumed to have access to the detector noise, as it arises locally at Bob's measurement stage. Therefore, when calculating Eve’s Holevo bound, we conservatively assume that she has access to a state where $\eta = 1$ and $v_{el} = 0$. \ozlem{Before Bob rescales his data by $N_d$, the covariance matrix describing Alice and Bob’s modes after Bob’s redisplacement takes the form~\cite{qi2016simultaneous,zaunders2024quantum,zaunders2025enhanced}
\begin{equation}
\label{eq:matrix_without_Nd}
\sigma_{AB}=\begin{pmatrix}
    V\mathbb{I}_2 & C_d\sigma_z\\
    C_d\sigma_z & V_{b_d}\mathbb{I}_2
\end{pmatrix},
\end{equation}
from which the channel transmittance $T$ and excess noise $\xi$ can be inferred prior to rescaling. In bounding Eve’s information, we therefore set $\eta = 1$ and $v_{el} = 0$, ensuring that all channel-induced errors are attributed to Eve while detector inefficiencies and electronic noise remain trusted and local to Bob. This distinction underlies the trusted-receiver model.} While these idealised parameters \ozlem{are used for} Eve’s potential information, the actual detector inefficiencies and noise cannot be removed from the data shared between Alice and Bob. As a result, the mutual information $I_{AB}$ is degraded by the additional loss and noise present in Bob’s detection. Let us assume that the state that Eve sees is
\begin{equation}
    \label{eq:eves_state}
    \sigma_{AB}'(\eta\!=\!1,\!v_{el}\!=\!0)\!=\!\begin{pmatrix}
    \Tilde{V}\mathbb{I}_2 & N_d'\Tilde{C}_d'\sigma_z \\
     N_d'\Tilde{C}_d'\sigma_z & \Tilde{V_b}'\mathbb{I}_2    
    \end{pmatrix}\!=\!\begin{pmatrix}
        \Tilde{a}\mathbb{I}_2 & \Tilde{c}'\sigma_z \\
        \Tilde{c}'\sigma_z & \Tilde{b}'\mathbb{I}_2    
    \end{pmatrix}\!.
\end{equation}
Eve's Holevo bound can be expressed as
\begin{equation}
    \label{eq:holevo_bound}
    I_{E}=S(\Tilde{\sigma}_{AB}')-S(\Tilde{\sigma}_{A|b}'),
\end{equation}
\begin{figure*}[t!]
\center{\includegraphics[scale=0.56]
        {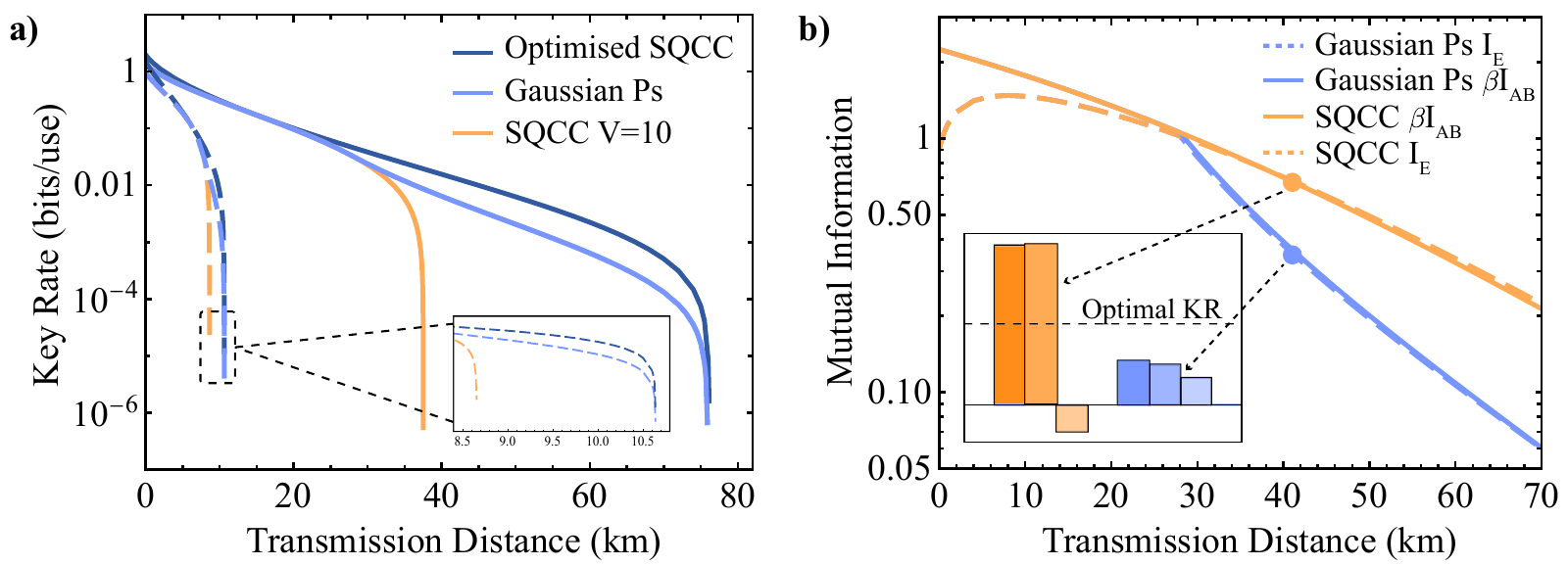}}
\caption{\label{fig:figure2}The asymptotic results assuming a fibre loss of 0.2~dB/km and thermal noise $\xi = 0.05$, expressed in shot-noise units (SNU). \textbf{(a)} The orange lines show the SQCC protocol with a fixed modulation variance of $V = 10$~SNU. The navy blue lines correspond to the SQCC protocol where the modulation variance is optimised at each distance, while the light-blue lines show SQCC with Gaussian post-selection with the filter gain optimised per distance. \ozlemREV{Solid lines correspond to the free-space model in a terrestrial setting, using a detector efficiency of $\eta = 0.95$ and detector noise of $v_{\mathrm{el}} = 0.01$~SNU, whereas dashed lines represent the fibre-based model with $\eta = 0.68$ and $v_{\mathrm{el}} = 0.05$~SNU.} Panel \textbf{(b)} \ozlemREV{illustrates the free-space model, where} solid lines represent the mutual information between Alice and Bob, while dashed lines show Eve’s Holevo information. The orange lines correspond to the SQCC protocol with $V = 10$, and the blue lines to the protocol with Gaussian post-selection. The dots indicate the point at $d = 41$~km, which is shown in detail in the panel. From left to right, we plot Alice and Bob’s mutual information, Eve’s information, and the key rate for both protocols: orange boxes show the original SQCC results, where the key rate is negative at this distance, while blue boxes show the SQCC protocol with Gaussian post-selection, where the key rate becomes positive with the optimal filter gain $g=0.25$. Inset: columns representing $\beta \mathrm{I_{AB}}$, $\mathrm{I_{E}}$ and $\mathrm{KR}$, respectively. $\mathrm{I_{AB}}$: Mutual information between Alice and Bob, $\mathrm{I_E}$: Eve's information and $\beta$: Reconciliation efficiency. \ozlem{For these simulations, the reconciliation efficiency is fixed at $\beta = 95\%$, while Bob’s classical bit-error rate, $\mathcal{W}$, varies with the channel parameters. The classical displacement is kept constant at $d=60$.} We assume the detection efficiency is $\eta=0.95$ and the detector noise is $v_{el}=0.01$~SNU (Refer to Appendix~\ref{sec:parameters_general} for the simulation parameters).
}
\end{figure*}
where $S(\Tilde{\sigma}_{AB}')$ and $S(\Tilde{\sigma}_{A|b}')$ represent the von Neumann entropy of Alice and Bob's average state and Alice's state conditioned on Bob's measurement outcomes $(x_b,p_b)$, respectively. Alice's conditional covariance matrix can be found from~\cite{laudenbach2018continuous}
\begin{equation}
    \label{eq:eves_conditional_state}
    \Tilde{\sigma}_{A|b}'=\sigma_{a}-\frac{1}{\Tilde{b}'+1}\sigma_c\sigma_c^T,
\end{equation}
where $\sigma_a=\Tilde{a}\mathbb{I}_2$ is Alice's covariance matrix and $\sigma_c=\Tilde{c}'\sigma_z$ is the covariance term between Alice and Bob given in Eq.~\ref{eq:eves_state}. The von Neumann entropies of both $\Tilde{\sigma}_{AB}'$ and $\Tilde{\sigma}_{A|b}'$ are calculated from their symplectic eigenvalues. The symplectic eigenvalues of $\Tilde{\sigma}_{AB}'$ are
\begin{equation}
    \label{eq:eigenvalues_sigma_ab}
    \lambda_{1,2}=\sqrt{\frac{1}{2}\big(\Delta\pm\sqrt{\Delta^2-4\mathrm{det(\Tilde{\sigma}_{AB}')}}\big)},
\end{equation}
where $\Delta=\mathrm{det}(\sigma_a)+\mathrm{det}(\sigma_b)+2\mathrm{det}(\sigma_c)$ and $\sigma_b=\Tilde{b}'\mathbb{I}_2$. The symplectic eigenvalues of $\Tilde{\sigma}_{A|b}'$ is obtained from
\begin{equation}
    \lambda=\sqrt{\mathrm{det}(\Tilde{\sigma}_{A|b}')}.
\end{equation}
The von Neumann entropy is computed from
\begin{equation}
    \label{eq:entropy_covariance_matrix}
    S(\sigma)=\sum_{i=1}^N g(\lambda_i),
\end{equation}
where
\begin{equation}
    \label{eq:g_function}
    g(x)\!=\!\bigg(\frac{x+1}{2}\bigg)\!\log_2\!\bigg(\frac{x+1}{2}\bigg)\!-\!\bigg(\frac{x-1}{2}\bigg)\!\log_2\!\bigg(\frac{x-1}{2}\bigg).
\end{equation}
We maximise the asymptotic key rate by optimising the filter gain applied by Alice, which is defined as
\begin{equation}
    \label{eq:optimal_asymptotic_key_rate}
    K^{\infty}_{opt}(\Tilde{\sigma}_{AB},\Tilde{\sigma}_{AB}')=\max_{g}\big[P_A\big(\beta I_{AB}(\Tilde{\sigma}_{AB})-I_E(\Tilde{\sigma}_{AB}')\big)\big]
\end{equation}
\ozlem{Note that Alice’s classical displacement $d$ is usually set according to a target quality-of-service requirement, typically by fixing the classical bit-error rate $\mathcal{W}$. In this case, $d$ must satisfy}
\begin{equation}
    d\geq 2\;\text{erfc}^{-1}(2\mathcal{W})\sqrt{\frac{V_b+1}{T}}.
\end{equation}
\ozlem{However, this approach presumes prior knowledge or reliable characterisation of the channel, which may not hold in fluctuating free-space or satellite links. In our analysis, we therefore depart from this assumption and instead take Alice’s displacement $d$ to be fixed.}

\subsection{Finite-Size Analysis}
As finite-size affects the estimation of the covariance matrix, the post-selected covariance matrix also scales with this estimation. Following the approach outlined in Ref.~\cite{zaunders2025enhanced}, the estimated covariance matrix is given by
\begin{equation}
    a_{max}=(1+\delta_{\mathrm{Var}})\Tilde{a},
\end{equation}
\begin{equation}
    b_{max}=(1+\delta_{\mathrm{Var}})\Tilde{b},
\end{equation}
\begin{equation}
    c_{min}=\bigg(1-2\sqrt{\frac{ab}{c^2}}\delta_{\mathrm{Cov}}\bigg)\Tilde{c},
\end{equation}
where
\begin{equation}
    \label{eq:delta_var}
    \delta_{\mathrm{Var}}=\bigg[2-A\bigg(\frac{\epsilon_{\mathrm{PE}}}{12}\bigg)\bigg]\bigg[1+\frac{240}{\epsilon_{\mathrm{PE}}}e^{-N/32}\bigg]-1,
\end{equation}
\begin{equation}
    \label{eq:delta_cov}
    \delta_{\mathrm{Cov}}=\frac{1}{2}\bigg[1-A\bigg(\frac{\epsilon_{\mathrm{PE}}}{12}\bigg)\bigg]+\bigg[1-A\bigg(\frac{\epsilon_{\mathrm{PE}}^2}{1296}\bigg)\bigg],
\end{equation}
and
\begin{equation}
    \label{eq:A_definition}
A(z)=2\mathrm{invcdf}_{\mathrm{Beta}[\frac{N}{2},\frac{N}{2}]}(z),
\end{equation}
\ozlemblue{where $\mathrm{invcdf_{Beta}}$ denotes the inverse cumulative distribution function of the Beta distribution.}
Therefore, the finite-size covariance matrix becomes
\begin{equation}
    \sigma_{AB}^{fs}=\begin{pmatrix}
        a_{max}\mathbb{I}_2 & c_{max}\sigma_z \\
         c_{max}\sigma_z & b_{max}\mathbb{I}_2
    \end{pmatrix}.
\end{equation}
Similar to the asymptotic case, we construct a separate covariance matrix between Alice and Bob, $\sigma_{AB}^{fs'}$, for estimating Eve's information under the assumption $\eta = 1$ and $v_{el} = 0$. The post-selected finite-size key rate is then given by
\begin{align}
    K^{fs}_{ps}&=P_Ap_F(I_{AB}(\sigma_{AB}^{fs})-I_E(\sigma_{AB}^{fs'}))-\sqrt{\frac{p_fP_A}{N}}\Delta_{\mathrm{AEP}} \nonumber \\
    &-\sqrt{\frac{p_fP_A\log_2(p_fP_AN)}{N}}\Delta_{\mathrm{ent}}\nonumber+\frac{\Delta_S}{N}+\frac{\Delta_H}{N},
\end{align}
where $P_A$ is the post-selection success probability defined in Eq.~\eqref{eq:alices_ps}. It can also be expressed as $P_A = N_{ps} / N$, where $N_{ps}$ is the number of symbols kept after post-selection, and $N$ is the total number of symbols used for parameter estimation.


\section{\label{sec:results}Results}
\begin{figure}[t]
\includegraphics[scale=0.58]{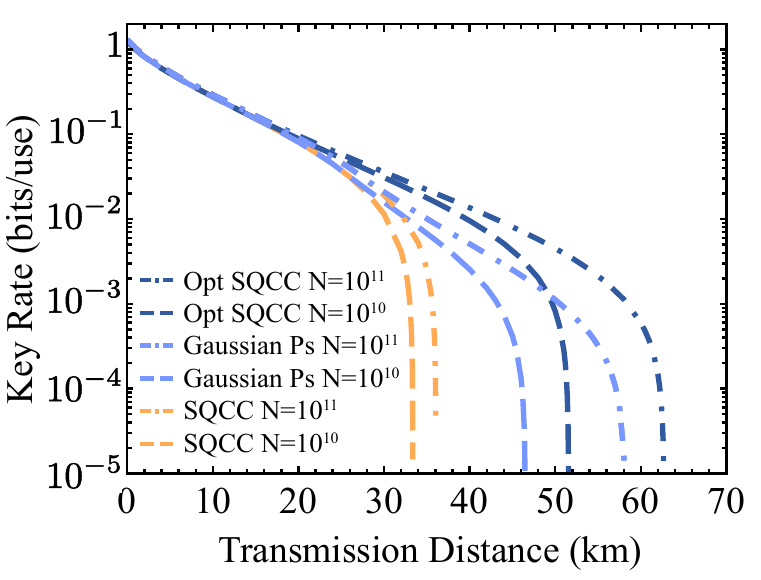}\hspace*{-0.1cm}
\caption{\label{fig:figure3}Finite-size key rate performance of the \ozlemREV{free-space SQCC} protocol over a thermal channel with excess noise $\xi=0.05$ and \ozlem{clasical displacement of $d=60$,} shown for different block sizes. Dashed lines represent a block size of $N=10^{10}$, while the dot-dashed lines correspond to a block size of $N=11$. The orange line shows the original SQCC protocol with $V=10$~SNU, the light blue line includes Gaussian post-selection, and the dark blue line shows the SQCC protcol with optimised variance. \ozlemREV{The detector efficiency and noise are set to $\eta=0.95$ and $v_{el}=0.01$~SNU, respectively.}
}
\end{figure}
Figure~\ref{fig:figure2} illustrates the performance of the SQCC protocol with and without post-selection in the asymptotic regime \ozlemREV{for both fibre and free-space based QKD}. As shown in Fig.~\ref{fig:figure2}(a), the original SQCC protocol with a fixed thermal variance of $V = 10$ fails to generate a key beyond $d = 37.5$~km \ozlemREV{for the free-space model in a terrestrial setting}. In contrast, the SQCC protocol with Gaussian post-selection achieves a notable improvement \ozlemREV{over the fixed-variance SQCC protocol}, extending the key distribution range by 38.5~km and enabling key generation up to $d = 76$~km. \ozlemREV{However, the improvement is modest for the fibre-based scenario, as detector efficiencies in fibre systems are generally lower than in free-space implementations. For example, with a detector efficiency of $\eta = 0.68$ and electronic noise of $v_{\mathrm{el}} = 0.05$~SNU, representative of current fibre-based homodyne detectors~\cite{jain2022practical, hajomer2024long}, the post-selected SQCC scheme increases the achievable distance by roughly $2$~km as shown in Fig.~\ref{fig:figure2}(a). This smaller improvement is due to the additional losses introduced by fibre coupling and imperfect mode matching. We note that the free-space parameters used in our simulations correspond to the favourable end of currently achievable detector performance; however, as the detector efficiency decreases and the electronic noise increases, the key rates reduce and the entire key-rate curve shifts towards shorter distances. In such cases, the Gaussian post-selection protocol should continue to provide an improvement over the fixed-variance SQCC scheme.}

It is important to note that the post-selection protocol performs below the SQCC protocol with optimised variance at each distance. This is due to its probabilistic nature, some data is discarded during post-selection, resulting in a penalty from the success probability. For instance, at $d = 41$~km, the post-selected key rate is scaled by a success probability of $P_s = 0.45$ for a filter gain of $g=0.25$. The strength of this protocol lies in its adaptability to unknown or time-varying channels. When channel parameters cannot be predicted in advance, post-selection provides a practical means of recovering near-optimal key rates. Additionally, since the post-selection filter does not correspond to a physical process, we have the freedom to tune the filter gain to maximise performance. In essence, the filter emulates the effect of preparing a state with the optimal variance that would have been chosen had the channel been known. For example, when the gain is set to $g = 0$, no filtering occurs and the variance remains unchanged. As the filter gain $g$ increases, more of the distribution is truncated, effectively reducing the variance. In the limit of large $g$, the post-selected states resemble vacuum states, as only inputs near the origin are kept.
\begin{figure*}[t!]
\center{\includegraphics[scale=0.42]
        {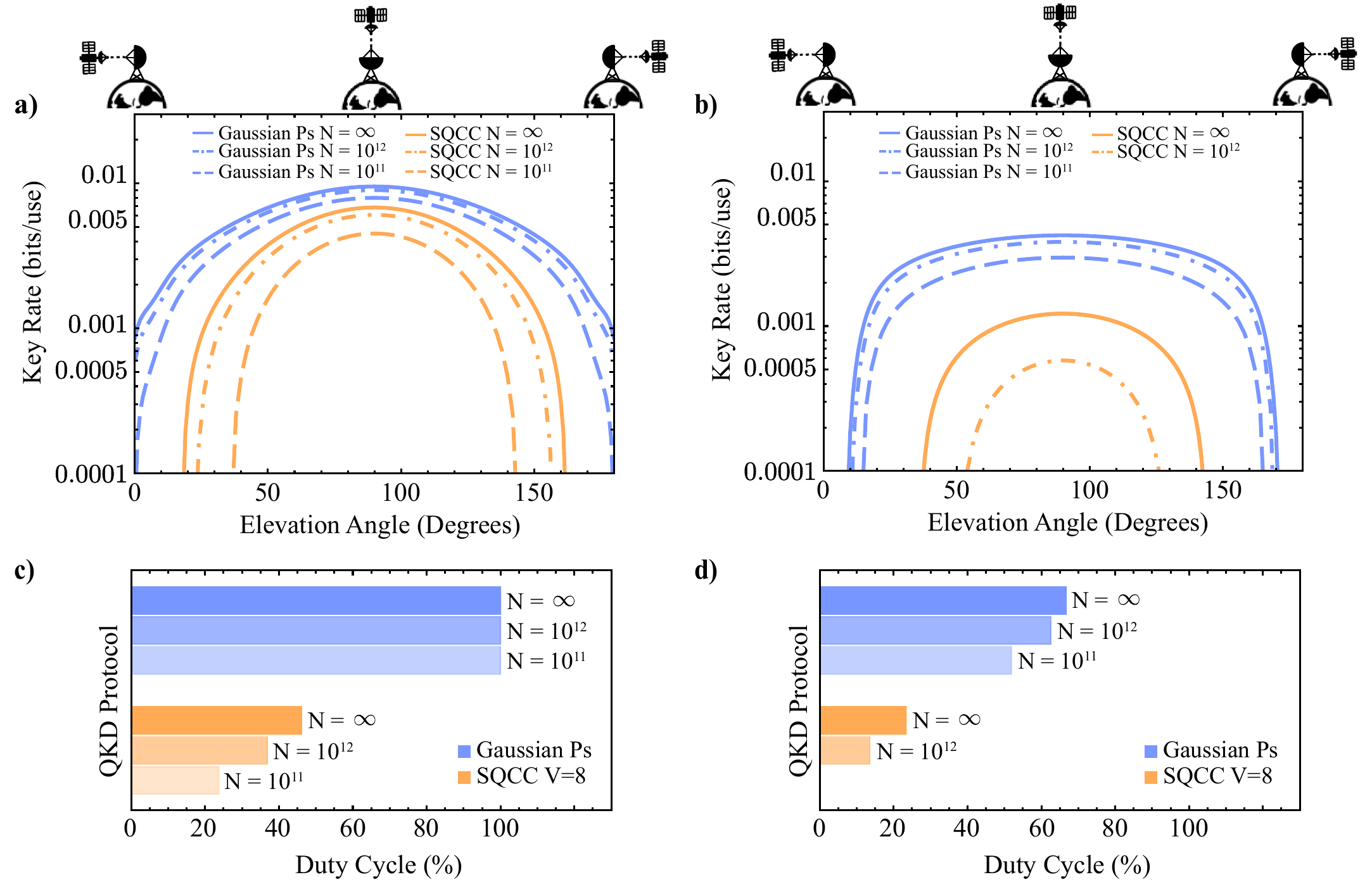}}
\caption{\label{fig:figure4}Key rate of satellite-to-ground communication as a function of the elevation angle, shown for both the original SQCC protocol and the SQCC protocol with Gaussian post-selection. The satellite is assumed to be in Low Earth Orbit at an altitude of 500~km, with the optical ground station located at sea level ($0$~km elevation). We adopt the satellite-to-ground channel model from Sayat \textit{et al.}~\cite{sayat2024satellite}, assuming a reconciliation efficiency of $\beta = 92\%$ (see Appendix~\ref{sec:parameter_satellite} for details). \textbf{(a)} The orange lines correspond to key rates using a fixed thermal variance of $V = 8$~SNU, while the blue lines show the performance of the SQCC protocol with Gaussian post-selection applied. Solid lines indicate the asymptotic key rates, while the dot-dashed and dashed lines represent finite-size results with block sizes of $N=10^{12}$ and $N=10^{11}$, respectively. The simulation assumes good weather conditions, with a visibility of $V_{visib} = 200$~km and low atmospheric turbulence characterised by $C_n^2 = 10^{-16}$~m$^{-2/3}$. At each elevation angle, the post-selection gain is optimised to maximise the key rate. \textbf{(b)} Key rates under adverse weather conditions, assuming a visibility of $V_{visib} = 20$~km and strong atmospheric turbulence with $C_n^2 = 10^{-13}$~m$^{-2/3}$. \textbf{(c)} Duty cycle comparison between the SQCC protocol with and without post-selection under good weather conditions, based on the key rate threshold of $10^{-4}$ from (a). \textbf{(d)} Duty cycle comparisonbetween the SQCC protocol with and without post-selection under adverse weather conditions, using the same key rate threshold of $10^{-4}$ from (b).
}
\end{figure*}

Figure~\ref{fig:figure2}(b) shows Alice and Bob’s mutual information alongside Eve’s Holevo bound as a function of distance for \ozlemREV{the free-space model.} Below $29$~km, the results of both the original SQCC protocol and the post-selected version coincide, as the fixed variance of $V = 10$~SNU remains optimal within this range. However, beyond this distance, the two curves diverge as post-selection takes effect. Both Alice and Bob’s mutual information and Eve’s Holevo information decrease compared to the original SQCC protocol. As previously discussed and shown in Fig.\ref{fig:figure2}(a), no key can be extracted beyond 37.5~km, since Eve’s information exceeds that of Alice and Bob, as also illustrated in Fig.~\ref{fig:figure2}(b). Post-selection mitigates this by reducing Eve’s information at a faster rate than the mutual information between Alice and Bob. Although both quantities decrease, this imbalance leads to a net positive key rate. This effect is illustrated in the inset of Fig.~\ref{fig:figure2}(b). At $41$~km, for instance, Eve’s information exceeds Alice and Bob’s, resulting in a negative key rate of $-5.64\times10^{-3}$~bits/use corresponding to an insecure regime. After applying post-selection, the key rate improves to $5.69\times10^{-3}$~bits/use. This is achieved by tuning the filter gain in Eq.\eqref{eq:alices_filter} to $g = 0.25$, which gives the maximum key rate at this distance.

Post-selection is also applied in the finite-size regime, with results shown in Fig.~\ref{fig:figure3} for different block sizes. When the variance of the SQCC protocol is fixed at $V=10$~SNU, the maximum transmission distances are 33~km and 36~km for block sizes of $N=10^{10}$ and $N=10^{11}$, respectively. By introducing Gaussian post-selection, where the filter gain is optimised at each distance, these limits extend to 46~km and 58~km for the same block sizes, resulting in improvements of $13$~km and $22$~km, respectively. Similar to the asymptotic case, the key rates achieved using Gaussian post-selection remain lower than those obtained with optimised variance due to the limited success probability of the post-selection. Nevertheless, it is promising that even in the finite-size regime, Gaussian post-selection still provides a noticeable performance benefit \ozlemREV{relative to the fixed-variance SQCC protocol.}

The enhancements to the SQCC protocol are not limited to terrestrial environments and can also be extended to satellite-to-ground links. Figure~\ref{fig:figure4}(a) and (b) illustrate how our protocol extends the communication window between a ground station and a satellite under different weather conditions, based on the satellite-to-ground channel model proposed by Sayat \textit{et al.}~\cite{sayat2024satellite}. Assuming a low-Earth orbit satellite at an altitude of 500~km, the orange lines correspond to the standard SQCC protocol with a variance of $V=8$, while the blue lines represent the performance of the SQCC protocol incorporating Alice’s Gaussian post-selection (See the Appendix~\ref{sec:parameter_satellite} for a detailed description of the simulation parameters). Under both good and bad weather conditions, the SQCC protocol with Gaussian post-selection outperforms the standard SQCC protocol with fixed modulation variance. Figure~\ref{fig:figure4}(a) shows that under good weather conditions, the standard SQCC protocol enables communication with the ground station over elevation angles ranging from $18^\circ$ to $162^\circ$ in the asymptotic regime. This range narrows to $24^\circ$–$156^\circ$ for $N = 10^{12}$ and further to $37^\circ-$$143^\circ$ for $N = 10^{11}$. In contrast, the protocol with Gaussian post-selection maintains secure key rates across the entire elevation range and offers up to a 40-fold improvement at the extreme angles where the standard SQCC protocol gives its lowest key rates. This is further reflected in the duty cycle plot in Fig.~\ref{fig:figure4}(c) where the protocol with Gaussian post-selection provides a longer active communication window, defined as the duration over which the key rate remains above $10^{-4}$ bits per use. Under good weather conditions, the standard SQCC protocol achieves a duty cycle between $24\%$ and $46\%$, corresponding to a communication window of 43 minutes to 1 hour and 23 minutes. In contrast, the post-selected protocol maintains a full 3-hour communication window with a $100\%$ duty cycle \ozlemREV{compared with the fixed-variance SQCC protocol}, even in the finite-size regime. 

In adverse weather conditions, as shown in Fig.~\ref{fig:figure4}(b), the standard SQCC protocol supports satellite-to-ground communication over elevation angles ranging from $37.5^\circ$ to $142.5^\circ$ in the asymptotic regime, and from $54^\circ$ to $126^\circ$ for a block size of $N = 10^{12}$, corresponding to duty cycles of $23\%$ and $13.5\%$, respectively. At a block size of $N = 10^{11}$, the standard SQCC protocol is unable to produce any secure key rates. 
In contrast, the post-selected protocol maintains a communication window at elevation angles from $9^\circ$ to $171^\circ$ for the asymptotic regime, while it reduces to $10.5^\circ-$$169.5^\circ$ and $15^\circ-$$165^\circ$ for a block size of $N = 10^{12}$ and $N = 10^{11}$, respectively. As a result, the communication window of the post-selected protocol ranges from 1 hours and 33 minutes to 2 hours, whereas the standard scheme supports only 24 to 42 minutes. This showcases the robustness of the Gaussian post-selected protocol, as the optimisation step for achieving the best key rates is performed in software, requiring no hardware modifications. This is particularly advantageous in free-space channels where conditions can fluctuate over time, making it impractical to rely on pre-characterised channel parameters for optimising the experiment before transmitting quantum states to the receiver. Therefore, by allowing real-time adaptability without hardware modifications, Gaussian post-selection improves the practical applicability of the SQCC protocol across a range of quantum communication platforms, including satellite and fibre links.

\section{\label{sec:discussion}Discussion}
SQCC enhances the practicality of CV-QKD protocols by multiplexing classical and quantum symbols onto a single optical pulse, making it particularly well-suited for platforms with limited channels or constrained resources. Despite its practical advantages, it faces the same challenges as standard CV-QKD protocols. 

While experimental parameters can be optimised to maximise key rates once the channel has been characterised, such optimisation is ineffective in fluctuating channels, where any estimated parameters quickly become outdated. To address this challenge, we introduced a Gaussian filtering process applied at Alice's station to the modulation data, after the quantum states have been sent and the channel has been estimated. This post-selection effectively emulates the optimisation of the modulation variance that would have been performed if Alice and Bob had prior knowledge of the channel parameters. This is particularly useful in fluctuating free-space channels, where channel estimation is possible but often unreliable due to variations and this method allows them to dynamically optimise the key rates.

The results presented in this paper demonstrate that the transmission distance of the SQCC protocol can be extended in both the asymptotic and finite-size regimes using Gaussian post-selection. This improvement holds across both fibre-based and free-space channels, with performance comparable to the SQCC protocol using optimised modulation variance. Future work could explore the performance of the post-selected SQCC protocol under more realistic fading channel models, such as those encountered in turbulent free-space links. In addition, investigating alternative post-selection techniques beyond Gaussian filtering may offer further improvements in key rate and robustness across varying channel conditions.

\section*{Acknowledgments}
The Australian Government supported this research through the Australian Research Council’s Linkage Projects funding scheme (Project No. LP200100601). The views expressed herein are those of the authors and are not necessarily those of the Australian Government or the Australian Research Council.
\vspace{-1em}

\appendix
\section{\label{sec:pessimistic_analysis}Security Analysis with Untrusted Receiver}
\ozlem{If the receiver detectors are treated as untrusted, Eve’s information is evaluated directly from the Alice and Bob's covariance matrix in Eq.~\eqref{eq:postselected_matrix}, using Gaussian extremality. However, this effectively attributes the additional loss and noise to Eve, causing both standard SQCC and Gaussian post-selected SQCC to perform worse than the trusted-receiver model. To achieve comparable results under this untrusted assumption, the receiver detectors would need to exhibit lower loss and noise.} 

\begin{figure*}[t]
\includegraphics[scale=0.55]{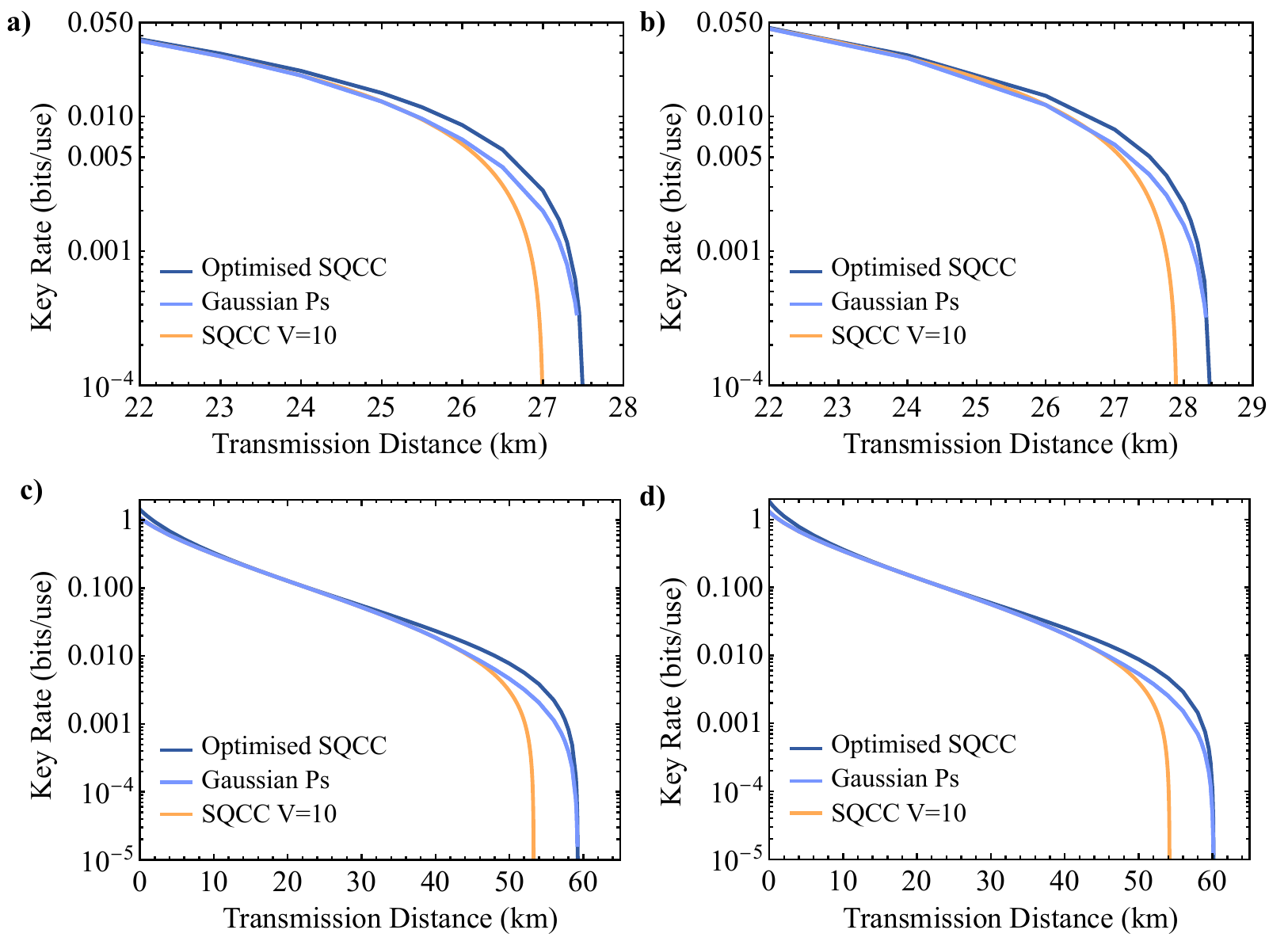}
\caption{\label{fig:figure5}Asymptotic key rates in the untrusted-detector scenario. Alice’s modulation variance is set to $V = 10$~SNU, with channel noise $\xi = 0.05$~SNU. \textbf{(a)} Detector efficiency $\eta = 95\%$, and electronic noise $v_{el} = 0.01$~SNU. \textbf{(b)} Detector efficiency $\eta = 99\%$, and electronic noise $v_{el} = 0.01$~SNU. \textbf{(c)} Detector efficiency $\eta = 95\%$, and electronic noise $v_{el} = 0.001$~SNU. \textbf{(d)} Detector efficiency $\eta = 99\%$, and electronic noise $v_{el} = 0.001$~SNU.
}
\end{figure*}

\ozlem{For example, using the same parameters as in Fig.~\ref{fig:figure2}, the standard SQCC protocol achieves secure keys up to $27$ km, while the optimised SQCC extends this range by only $0.5$~km. Likewise, applying Gaussian postselection provides a similar improvement, as illustrated in Fig.~\ref{fig:figure5}(a). When the detector efficiency is increased to $\eta = 0.99$ while maintaining $v_{el} = 0.01$~SNU, both our protocol and the standard SQCC gain an additional $1$~km in secure distance with the Gaussian post-selection showing $0.5$~km improvement in comparison to the standard SQCC scheme. Similar performance to the trusted scenario can be obtained when the detector efficiency is kept at $\eta = 0.95$ and the electronic noise reduced to $v_{el} = 0.001$~SNU as shown in Fig.~\ref{fig:figure5}(c). These parameter values are well within reach, as state-of-the-art CV detectors usually achieve efficiencies above $99\%$. The benefits are likely to be even more pronounced in regimes where Alice's initial variance is higher or lower reconciliation efficiencies are used. However, further increasing the efficiency to $\eta = 0.99$ does not result in noticeable gains in key rates or transmission distance, as shown in Fig.~\ref{fig:figure5}(d). In both panels (c) and (d), the Gaussian post-selected protocol consistently provides an overall enhancement of about $6$~km. This suggests that, in the untrusted model, electronic noise plays the dominant role in limiting transmission distance, whereas detector efficiency has a comparatively minor effect.}

\section{\label{sec:parameters_general}Simulation Parameters for Terrestrial Communications}
In this section, we summarise the parameters used to produces Figs.~\ref{fig:figure2} and Figs.~\ref{fig:figure3} as shown in detail in Table~\ref{tab:table1}.
\begin{table}[H]
\caption{Parameters used for Figs.~\ref{fig:figure2} and ~\ref{fig:figure3}.}
    \centering
    \begin{tabular}{|c|c|}
    \hline
    \textbf{Parameter} & \textbf{Value} \\
    \hline
    \makecell{Modulation Variance \\ $V_{mod}$~(SNU)} & 7 \\
    \hline
    \makecell{Excess Noise \\ $\xi$~(SNU)} & 0.05 \\
    \hline
    \makecell{Detector Efficiency \\ $\eta$~$(\%)$} & \makecell{95$\%$~(free-space)\\ 68$\%$~(fibre-based)} \\
    \hline
    \makecell{Detector Noise \\ $v_{el}$~(SNU)} & \makecell{0.01~(free-space) \\ 0.05~(fibre-based)} \\
    \hline
    \makecell{Classical Displacement \\ $d$~(SNU)} & 60 \\
    \hline
    \makecell{Reconciliation Efficiency \\ $\beta$ (\%)} & 95$\%$ \\
    \hline
    \makecell{No. of Discretisation Bits for \\ ADC of Gaussian Data, $d_{rx}$} & 6 \\
    \hline
    \makecell{Success Probability of Error \\ Correction per Frame, $p_f$} & 0.9964~\cite{jain2022practical} \\
    \hline
    \makecell{Confidence Level in Estimating \\ the Entropy, $\epsilon_{ent}$} & $10^{-10}$ \\
    \hline
    \makecell{Min-entropy Smoothing \\ Parameter, $\epsilon_{s}$} & $10^{-10}$ \\
    \hline
    \makecell{Left-over Hash Lemma \\ Confidence Parameter, $\epsilon_{h}$} & $10^{-10}$ \\
    \hline
    \makecell{Confidence Level of \\ Parameter Estimation, $\epsilon_{\text{PE}}$} & $10^{-10}$ \\
    \hline
    \end{tabular}
    \label{tab:table1}
\end{table}

\section{\label{sec:parameter_satellite}Simulation Parameters for Satellite-to-Ground Communication}
In this section, we provide the parameters used for the satellite-to-ground communication scenario presented in Fig.~\ref{fig:figure4}, summarised in Table~\ref{tab:table2}. The simulation model is based on Refs.~\cite{sayat2024satellite, erkilic2025enhanced}. \ozlemREV{We note that atmospheric turbulence primarily impacts the effective channel transmittance rather than the intrinsic detector efficiency, and is therefore modelled as part of the free-space channel rather than the receiver.}

\begin{table}[H]
\caption{\label{tab:free_space_parameters_table}
Parameters used in the satellite-to-ground communication model shown in Fig.~\ref{fig:figure4}.}
\centering
\begin{tabular}{|c|c|}
\hline
\textbf{Parameter} & \textbf{Value} \\
\hline
\makecell[c]{Radius of the Earth \\ $R_E$ (km)} & 6371 \\
\hline
\makecell[c]{Satellite Altitude at \\ Zenith $L_{\text{zen}}$ (km)} & 500 \\
\hline
\makecell[c]{Optical Ground Station \\ Elevation $L_{\text{OGS}}$ (km)} & 0 \\
\hline
\makecell[c]{Transmitter Aperture \\ Diameter $D_t$ (m)} & 0.3 \\
\hline
\makecell[c]{Receiver Aperture \\ Diameter $D_r$ (m)} & 1 \\
\hline
\makecell[c]{Transmitter Optics \\ Efficiency $T_t$} & 0.95 \\
\hline
\makecell[c]{Receiver Optics \\ Efficiency $T_r$} & 0.95 \\
\hline
\makecell[c]{Pointing Loss Efficiency \\ $L_p$} & 0.1 \\
\hline
\makecell[c]{Atmosphere Thickness \\ $L_{\text{atm}}$ (km)} & 20 \\
\hline
\makecell[c]{Visibility \\ $V$ (km)} & \makecell{20 (bad weather) \\ 200 (good weather)} \\
\hline
\makecell[c]{Refractive Index Structure \\ Parameter $C_n^2$ (m$^{-2/3}$)} & \makecell{$10^{-13}$ (bad weather) \\ $10^{-16}$ (good weather)} \\
\hline
\makecell[c]{Probability Threshold \\ $p_{\text{th}}$} & $10^{-6}$ \\
\hline
\makecell[c]{Wavelength \\ $\lambda$ (nm)} & 1550 \\
\hline
\makecell[c]{Modulation Variance \\ $V_{\text{mod}}$ (SNU)} & 7 \\
\hline
\makecell[c]{Detector Efficiency \\ $\eta$ (\%)} & 98.5\% \\
\hline
\makecell[c]{Detector Noise \\ $v_{el}$ (SNU)} & 0.01 \\
\hline
\makecell[c]{Excess Channel Noise \\ $\epsilon_{\text{ch}}$ (SNU)} & 0.02 \\
\hline
\makecell[c]{Detection Type} & Heterodyne \\
\hline
\makecell[c]{Reconciliation Efficiency \\ $\beta$ (\%)} & 92\% \\
\hline
\makecell[c]{Classical Displacement \\ $d$~(SNU)} & 50 \\
\hline
\end{tabular}
\label{tab:table2}
\end{table}

\bibliography{apssamp}

@PREAMBLE{
 "\providecommand{\noopsort}[1]{}" 
 # "\providecommand{\singleletter}[1]{#1}%" 
}

@article{pirandola2020advances,
  title={Advances in quantum cryptography},
  author={Pirandola, Stefano and Andersen, Ulrik L and Banchi, Leonardo and Berta, Mario and Bunandar, Darius and Colbeck, Roger and Englund, Dirk and Gehring, Tobias and Lupo, Cosmo and Ottaviani, Carlo and others},
  journal={Adv. Opt. Photonics},
  volume={12},
  number={4},
  pages={1012--1236},
  year={2020},
  publisher={Optical Society of America}
}

@article{ekert2014,
  title={The ultimate physical limits of privacy},
  author={Ekert, Artur and Renner, Renato},
  journal={Nature},
  volume={507},
  number={7493},
  pages={443--447},
  year={2014},
  publisher={Nature Publishing Group}
}

@article{gisin2002,
  title = {Quantum cryptography},
  author = {Gisin, Nicolas and Ribordy, Gr\'egoire and Tittel, Wolfgang and Zbinden, Hugo},
  journal = {Rev. Mod. Phys.},
  volume = {74},
  issue = {1},
  pages = {145--195},
  numpages = {0},
  year = {2002},
  month = {Mar},
  publisher = {American Physical Society},
  doi = {10.1103/RevModPhys.74.145}
}

@article{ralph1999,
  title = {Continuous variable quantum cryptography},
  author = {Ralph, T. C.},
  journal = {Phys. Rev. A},
  volume = {61},
  issue = {1},
  pages = {010303},
  numpages = {4},
  year = {1999},
  month = {Dec},
  publisher = {American Physical Society},
  doi = {10.1103/PhysRevA.61.010303}
}

@article{hillery2000,
  title = {Quantum cryptography with squeezed states},
  author = {Hillery, Mark},
  journal = {Phys. Rev. A},
  volume = {61},
  issue = {2},
  pages = {022309},
  numpages = {8},
  year = {2000},
  month = {Jan},
  publisher = {American Physical Society},
  doi = {10.1103/PhysRevA.61.022309}
}

@article{garcia2007quantum,
  title={Quantum information with optical continuous variables: from Bell tests to key distribution},
  author={Garcia-Patron Sanchez, Raul},
  year={2007},
  publisher={Universit{\'e} libre de Bruxelles}
}

@article{xiang2010heralded,
  title={Heralded noiseless linear amplification and distillation of entanglement},
  author={Xiang, Guo-Yong and Ralph, Timothy C and Lund, Austin P and Walk, Nathan and Pryde, Geoff J},
  journal={Nat. Photonics},
  volume={4},
  number={5},
  pages={316--319},
  year={2010},
  publisher={Nature Publishing Group UK London}
}

@article{huang2013performance,
  title={Performance improvement of continuous-variable quantum key distribution via photon subtraction},
  author={Huang, Peng and He, Guangqiang and Fang, Jian and Zeng, Guihua},
  journal={Phys. Rev. A},
  volume={87},
  number={1},
  pages={012317},
  year={2013},
  publisher={APS}
}

@article{ye2019improvement,
  title={Improvement of self-referenced continuous-variable quantum key distribution with quantum photon catalysis},
  author={Ye, Wei and Zhong, Hai and Liao, Qin and Huang, Duan and Hu, Liyun and Guo, Ying},
  journal={Opt. Express},
  volume={27},
  number={12},
  pages={17186--17198},
  year={2019},
  publisher={Optica Publishing Group}
}

@article{hu2021performance,
  title={Performance improvement of unidimensional continuous-variable quantum key distribution using zero-photon quantum catalysis},
  author={Hu, Junkai and Liao, Qin and Mao, Yun and Guo, Ying},
  journal={Quantum Inf. Process.},
  volume={20},
  pages={1--20},
  year={2021},
  publisher={Springer}
}

@article{hu2020continuous,
  title={Continuous-variable quantum key distribution with non-Gaussian operations},
  author={Hu, Liyun and Al-Amri, M and Liao, Zeyang and Zubairy, MS},
  journal={Phys. Rev. A},
  volume={102},
  number={1},
  pages={012608},
  year={2020},
  publisher={APS}
}

@article{winnel2020generalized,
  title={Generalized quantum scissors for noiseless linear amplification},
  author={Winnel, Matthew S and Hosseinidehaj, Nedasadat and Ralph, Timothy C},
  journal={Phys. Rev. A},
  volume={102},
  number={6},
  pages={063715},
  year={2020},
  publisher={APS}
}

@article{walk2013security,
  title={Security of continuous-variable quantum cryptography with Gaussian postselection},
  author={Walk, Nathan and Ralph, Timothy C and Symul, Thomas and Lam, Ping Koy},
  journal={Phys. Rev. A},
  volume={87},
  number={2},
  pages={020303},
  year={2013},
  publisher={APS}
}

@article{fiuravsek2012gaussian,
  title={Gaussian postselection and virtual noiseless amplification in continuous-variable quantum key distribution},
  author={Fiur{\'a}{\v{s}}ek, Jarom{\'\i}r and Cerf, Nicolas J},
  journal={Phys. Rev. A},
  volume={86},
  number={6},
  pages={060302},
  year={2012},
  publisher={APS}
}

@article{chrzanowski2014measurement,
  title={Measurement-based noiseless linear amplification for quantum communication},
  author={Chrzanowski, Helen M and Walk, Nathan and Assad, Syed M and Janousek, Jiri and Hosseini, Sara and Ralph, Timothy C and Symul, Thomas and Lam, Ping Koy},
  journal={Nat. Photonics},
  volume={8},
  number={4},
  pages={333--338},
  year={2014},
  publisher={Nature Publishing Group UK London}
}

@article{zhao2017characterization,
  title={Characterization of a measurement-based noiseless linear amplifier and its applications},
  author={Zhao, Jie and Haw, Jing Yan and Symul, Thomas and Lam, Ping Koy and Assad, Syed M},
  journal={Phys. Rev. A},
  volume={96},
  number={1},
  pages={012319},
  year={2017},
  publisher={APS}
}

@article{hosseinidehaj2020finite,
  title={Finite-size effects in continuous-variable quantum key distribution with Gaussian postselection},
  author={Hosseinidehaj, Nedasadat and Lance, Andrew M and Symul, Thomas and Walk, Nathan and Ralph, Timothy C},
  journal={Phys. Rev. A},
  volume={101},
  number={5},
  pages={052335},
  year={2020},
  publisher={APS}
}

@article{zhong2020virtual,
  title={Virtual zero-photon catalysis for improving continuous-variable quantum key distribution via Gaussian post-selection},
  author={Zhong, Hai and Guo, Ying and Mao, Yun and Ye, Wei and Huang, Duan},
  journal={Sci. Rep.},
  volume={10},
  number={1},
  pages={17526},
  year={2020},
  publisher={Nature Publishing Group UK London}
}

@article{li2016non,
  title={Non-Gaussian postselection and virtual photon subtraction in continuous-variable quantum key distribution},
  author={Li, Zhengyu and Zhang, Yichen and Wang, Xiangyu and Xu, Bingjie and Peng, Xiang and Guo, Hong},
  journal={Phys. Rev. A},
  volume={93},
  number={1},
  pages={012310},
  year={2016},
  publisher={APS}
}

@article{zhong2018self,
  title={Enhancing of self-referenced continuous-variable quantum key distribution with virtual photon subtraction},
  author={Zhong, Hai and Wang, Yijun and Wang, Xudong and Liao, Qin and Wu, Xiaodong and Guo, Ying},
  journal={Entropy},
  volume={20},
  number={8},
  pages={578},
  year={2018},
  publisher={MDPI}
}

@article{wolf2006extremality,
  title={Extremality of Gaussian quantum states},
  author={Wolf, Michael M and Giedke, Geza and Cirac, J Ignacio},
  journal={Phys. Rev. Lett.},
  volume={96},
  number={8},
  pages={080502},
  year={2006},
  publisher={APS}
}

@article{garcia2006unconditional,
  title={Unconditional Optimality of Gaussian Attacks against Continuous-Variable Quantum Key Distribution},
  author={Garc{\'\i}a-Patr{\'o}n, Ra{\'u}l and Cerf, Nicolas J},
  journal={Phys. Rev. Lett.},
  volume={97},
  number={19},
  pages={190503},
  year={2006},
  publisher={APS}
}

@article{grosshans2002continuous,
  title={Continuous variable quantum cryptography using coherent states},
  author={Grosshans, Fr{\'e}d{\'e}ric and Grangier, Philippe},
  journal={Phys. Rev. Lett.},
  volume={88},
  number={5},
  pages={057902},
  year={2002},
  publisher={APS}
}

@article{lance2005no,
  title={No-switching quantum key distribution using broadband modulated coherent light},
  author={Lance, Andrew M and Symul, Thomas and Sharma, Vikram and Weedbrook, Christian and Ralph, Timothy C and Lam, Ping Koy},
  journal={Phys. Rev. Lett.},
  volume={95},
  number={18},
  pages={180503},
  year={2005},
  publisher={APS}
}

@article{hajomer2024long,
  title={Long-distance continuous-variable quantum key distribution over 100-km fiber with local local oscillator},
  author={Hajomer, Adnan AE and Derkach, Ivan and Jain, Nitin and Chin, Hou-Man and Andersen, Ulrik L and Gehring, Tobias},
  journal={Sci. Adv.},
  volume={10},
  number={1},
  pages={eadi9474},
  year={2024},
  publisher={American Association for the Advancement of Science}
}

@article{weedbrook2012gaussian,
  title={Gaussian quantum information},
  author={Weedbrook, Christian and Pirandola, Stefano and Garc{\'\i}a-Patr{\'o}n, Ra{\'u}l and Cerf, Nicolas J and Ralph, Timothy C and Shapiro, Jeffrey H and Lloyd, Seth},
  journal={Rev. Mod. Phys.},
  volume={84},
  number={2},
  pages={621--669},
  year={2012},
  publisher={APS}
}

@article{grosshans2003virtual,
  title={Virtual entanglement and reconciliation protocols for quantum cryptography with continuous variables},
  author={Grosshans, Fr{\'e}d{\'e}ric and Cerf, Nicolas J and Wenger, J{\'e}r{\^o}me and Tualle-Brouri, Rosa and Grangier, Ph},
  journal={arXiv preprint quant-ph/0306141},
  year={2003}
}

@article{holevo1998capacity,
  title={The capacity of the quantum channel with general signal states},
  author={Holevo, Alexander S},
  journal={IEEE Trans. Inf. Theory},
  volume={44},
  number={1},
  pages={269--273},
  year={1998},
  publisher={IEEE}
}

@article{laudenbach2018continuous,
  title={Continuous-variable quantum key distribution with Gaussian modulation—the theory of practical implementations},
  author={Laudenbach, Fabian and Pacher, Christoph and Fung, Chi-Hang Fred and Poppe, Andreas and Peev, Momtchil and Schrenk, Bernhard and Hentschel, Michael and Walther, Philip and H{\"u}bel, Hannes},
  journal={Adv. Quantum Technol.},
  volume={1},
  number={1},
  pages={1800011},
  year={2018},
  publisher={Wiley Online Library}
}

@article{wang201525,
  title={25 MHz clock continuous-variable quantum key distribution system over 50 km fiber channel},
  author={Wang, Chao and Huang, Duan and Huang, Peng and Lin, Dakai and Peng, Jinye and Zeng, Guihua},
  journal={Sci. Rep.},
  volume={5},
  number={1},
  pages={14607},
  year={2015},
  publisher={Nature Publishing Group UK London}
}

@article{sayat2024satellite,
  title={Satellite-to-ground continuous variable quantum key distribution: The Gaussian and discrete modulated protocols in low earth orbit},
  author={Sayat, Mikhael and Shajilal, Biveen and Kish, Sebastian P and Assad, Syed M and Symul, Thomas and Lam, Ping Koy and Rattenbury, Nicholas and Cater, John},
  journal={IEEE Trans. Commun.},
  year={2024},
  publisher={IEEE}
}

@article{grosshans2003quantum,
  title={Quantum key distribution using gaussian-modulated coherent states},
  author={Grosshans, Fr{\'e}d{\'e}ric and Van Assche, Gilles and Wenger, J{\'e}r{\^o}me and Brouri, Rosa and Cerf, Nicolas J and Grangier, Philippe},
  journal={Nature},
  volume={421},
  number={6920},
  pages={238--241},
  year={2003},
  publisher={Nature Publishing Group UK London}
}

@article{jouguet2013experimental,
  title={Experimental demonstration of long-distance continuous-variable quantum key distribution},
  author={Jouguet, Paul and Kunz-Jacques, S{\'e}bastien and Leverrier, Anthony and Grangier, Philippe and Diamanti, Eleni},
  journal={Nat. Photonics},
  volume={7},
  number={5},
  pages={378--381},
  year={2013},
  publisher={Nature Publishing Group UK London}
}

@article{jain2022practical,
  title={Practical continuous-variable quantum key distribution with composable security},
  author={Jain, Nitin and Chin, Hou-Man and Mani, Hossein and Lupo, Cosmo and Nikolic, Dino Solar and Kordts, Arne and Pirandola, Stefano and Pedersen, Thomas Brochmann and Kolb, Matthias and {\"O}mer, Bernhard and others},
  journal={Nat. Commun.},
  volume={13},
  number={1},
  pages={4740},
  year={2022},
  publisher={Nature Publishing Group UK London}
}

@article{madsen2012continuous,
  title={Continuous variable quantum key distribution with modulated entangled states},
  author={Madsen, Lars S and Usenko, Vladyslav C and Lassen, Mikael and Filip, Radim and Andersen, Ulrik L},
  journal={Nat. Commun.},
  volume={3},
  number={1},
  pages={1083},
  year={2012},
  publisher={Nature Publishing Group UK London}
}

@article{lodewyck2007quantum,
  title={Quantum key distribution over 25 km with an all-fiber continuous-variable system},
  author={Lodewyck, J{\'e}r{\^o}me and Bloch, Matthieu and Garc{\'\i}a-Patr{\'o}n, Ra{\'u}l and Fossier, Simon and Karpov, Evgueni and Diamanti, Eleni and Debuisschert, Thierry and Cerf, Nicolas J and Tualle-Brouri, Rosa and McLaughlin, Steven W and others},
  journal={Phys. Rev. A},
  volume={76},
  number={4},
  pages={042305},
  year={2007},
  publisher={APS}
}

@article{weedbrook2004quantum,
  title={Quantum cryptography without switching},
  author={Weedbrook, Christian and Lance, Andrew M and Bowen, Warwick P and Symul, Thomas and Ralph, Timothy C and Lam, Ping Koy},
  journal={Phys. Rev. Lett.},
  volume={93},
  number={17},
  pages={170504},
  year={2004},
  publisher={APS}
}

@article{zhao2023enhancing,
  title={Enhancing quantum teleportation efficacy with noiseless linear amplification},
  author={Zhao, Jie and Jeng, Hao and Conlon, Lorc{\'a}n O and Tserkis, Spyros and Shajilal, Biveen and Liu, Kui and Ralph, Timothy C and Assad, Syed M and Lam, Ping Koy},
  journal={Nat. Commun.},
  volume={14},
  number={1},
  pages={4745},
  year={2023},
  publisher={Nature Publishing Group UK London}
}

@article{shajilal2024improving,
  title={Improving Gaussian channel simulation using non-unity gain heralded quantum teleportation},
  author={Shajilal, Biveen and Conlon, Lorc{\'a}n O and Walsh, Angus and Tserkis, Spyros and Zhao, Jie and Janousek, Jiri and Assad, Syed and Lam, Ping Koy},
  journal={arXiv preprint arXiv:2408.08667},
  year={2024}
}

@article{bencheikh2001quantum,
  title={Quantum key distribution with continuous variables},
  author={Bencheikh, Kamel and Symul, Th and Jankovic, A and Levenson, Juan-Ariel},
  journal={J. Mod. Opt.},
  volume={48},
  number={13},
  pages={1903--1920},
  year={2001},
  publisher={Taylor \& Francis}
}

@article{usenko2025continuous,
  title={Continuous-variable quantum communication},
  author={Usenko, Vladyslav C and Ac{\'\i}n, Antonio and All{\'e}aume, Romain and Andersen, Ulrik L and Diamanti, Eleni and Gehring, Tobias and Hajomer, Adnan AE and Kanitschar, Florian and Pacher, Christoph and Pirandola, Stefano and others},
  journal={arXiv preprint arXiv:2501.12801},
  year={2025}
}

@article{qi2016simultaneous,
  title={Simultaneous classical communication and quantum key distribution using continuous variables},
  author={Qi, Bing},
  journal={Physical Review A},
  volume={94},
  number={4},
  pages={042340},
  year={2016},
  publisher={APS}
}

@inproceedings{zaunders2024quantum,
  title={Quantum-Amplified Simultaneous Quantum-Classical Communications},
  author={Zaunders, Nicholas and Wang, Ziqing and Ralph, Timothy C and Aguinaldo, Ryan and Malaney, Robert},
  booktitle={2024 International Conference on Quantum Communications, Networking, and Computing (QCNC)},
  pages={160--167},
  year={2024},
  organization={IEEE}
}

@article{zaunders2025enhanced,
  title={Enhanced Simultaneous Quantum-Classical Communications Under Composable Security},
  author={Zaunders, Nicholas and Wang, Ziqing and Malaney, Robert and Aguinaldo, Ryan and Ralph, Timothy C},
  journal={arXiv preprint arXiv:2505.03145},
  year={2025}
}

@article{qi2018noise,
  title={Noise analysis of simultaneous quantum key distribution and classical communication scheme using a true local oscillator},
  author={Qi, Bing and Lim, Charles Ci Wen},
  journal={Physical Review Applied},
  volume={9},
  number={5},
  pages={054008},
  year={2018},
  publisher={APS}
}

@article{pan2020simultaneous,
  title={Simultaneous two-way classical communication and measurement-device-independent quantum key distribution with coherent states},
  author={Pan, Dong and Ng, Soon Xin and Ruan, Dong and Yin, Liuguo and Long, Guilu and Hanzo, Lajos},
  journal={Physical Review A},
  volume={101},
  number={1},
  pages={012343},
  year={2020},
  publisher={APS}
}

@article{winnel2024classical,
  title={Classical-quantum dual encoding for laser communications in space},
  author={Winnel, Matthew S and Wang, Ziqing and Malaney, Robert and Aguinaldo, Ryan and Green, Jonathan and Ralph, Timothy C},
  journal={New Journal of Physics},
  volume={26},
  number={3},
  pages={033012},
  year={2024},
  publisher={IOP Publishing}
}

@article{erkilic2025enhanced,
  title={Enhanced continuous-variable quantum key distribution protocol via adaptive signal processing},
  author={Erk{\i}l{\i}{\c{c}}, {\"O}zlem and Shajilal, Biveen and Conlon, Lorc{\'a}n O and Walsh, Angus and Das, Aritra and Kish, Sebastian and Symul, Thomas and Lam, Ping Koy and Assad, Syed M and Zhao, Jie},
  journal={Commun. Phys.},
  volume={8},
  number={1},
  pages={406},
  year={2025},
  publisher={Nature Publishing Group UK London}
}

@article{jeng2025entanglement,
  title={Entanglement-based quantum key distribution with non-Gaussian continuous variables},
  author={Jeng, Hao and Lam, Ping Koy and Assad, Syed M},
  journal={arXiv preprint arXiv:2507.18000},
  year={2025}
}

@article{chen2021continuous,
  title={Continuous-variable quantum key distribution based on photon addition operation},
  author={Chen, Xiao-Ting and Zhang, Lu-Ping and Chang, Shou-Kang and Zhang, Huan and Hu, Li-Yun},
  journal={Chinese Physics B},
  volume={30},
  number={6},
  pages={060304},
  year={2021},
  publisher={IOP Publishing}
}

@article{sasaki2011field,
  title={Field test of quantum key distribution in the Tokyo QKD Network},
  author={Sasaki, Masahide and Fujiwara, Mikio and Ishizuka, H and Klaus, W and Wakui, K and Takeoka, M and Miki, S and Yamashita, T and Wang, Z and Tanaka, A and others},
  journal={Opt. Express},
  volume={19},
  number={11},
  pages={10387--10409},
  year={2011},
  publisher={Optical Society of America}
}

@article{chen2021implementation,
  title={Implementation of a 46-node quantum metropolitan area network},
  author={Chen, Teng-Yun and Jiang, Xiao and Tang, Shi-Biao and Zhou, Lei and Yuan, Xiao and Zhou, Hongyi and Wang, Jian and Liu, Yang and Chen, Luo-Kan and Liu, Wei-Yue and others},
  journal={npj Quantum Info.},
  volume={7},
  number={1},
  pages={134},
  year={2021},
  publisher={Nature Publishing Group UK London}
}

@article{chen2010metropolitan,
  title={Metropolitan all-pass and inter-city quantum communication network},
  author={Chen, Teng-Yun and Wang, Jian and Liang, Hao and Liu, Wei-Yue and Liu, Yang and Jiang, Xiao and Wang, Yuan and Wan, Xu and Cai, Wen-Qi and Ju, Lei and others},
  journal={Opt. Express},
  volume={18},
  number={26},
  pages={27217--27225},
  year={2010},
  publisher={Optical Society of America}
}

@article{peev2009secoqc,
  title={The SECOQC quantum key distribution network in Vienna},
  author={Peev, Momtchil and Pacher, Christoph and All{\'e}aume, Romain and Barreiro, Claudio and Bouda, Jan and Boxleitner, Winfried and Debuisschert, Thierry and Diamanti, Eleni and Dianati, Mehrdad and Dynes, James F and others},
  journal={New J. Phys},
  volume={11},
  number={7},
  pages={075001},
  year={2009},
  publisher={IOP Publishing}
}

@article{pirandola2013high,
  title={High-rate quantum cryptography in untrusted networks},
  author={Pirandola, Stefano and Ottaviani, Carlo and Spedalieri, Gaetana and Weedbrook, Christian and Braunstein, Samuel L and Lloyd, Seth and Gehring, Tobias and Jacobsen, Christian S and Andersen, Ulrik L},
  journal={arXiv preprint arXiv:1312.4104},
  year={2013}
}

@article{hajomer2025high,
  title={High-rate continuous-variable measurement device-independent quantum key distribution with finite-size security},
  author={Hajomer, Adnan AE and Andersen, Ulrik L and Gehring, Tobias},
  journal={Quantum Science and Technology},
  volume={10},
  number={2},
  pages={025032},
  year={2025},
  publisher={IOP Publishing}
}

@article{chen2021integrated,
  title={An integrated space-to-ground quantum communication network over 4,600 kilometres},
  author={Chen, Yu-Ao and Zhang, Qiang and Chen, Teng-Yun and Cai, Wen-Qi and Liao, Sheng-Kai and Zhang, Jun and Chen, Kai and Yin, Juan and Ren, Ji-Gang and Chen, Zhu and others},
  journal={Nature},
  volume={589},
  number={7841},
  pages={214--219},
  year={2021},
  publisher={Nature Publishing Group UK London}
}

@article{chapuran2009optical,
  title={Optical networking for quantum key distribution and quantum communications},
  author={Chapuran, TE and Toliver, P and Peters, NA and Jackel, J and Goodman, MS and Runser, RJ and McNown, SR and Dallmann, N and Hughes, RJ and McCabe, KP and others},
  journal={New J. Phys},
  volume={11},
  number={10},
  pages={105001},
  year={2009},
  publisher={IOP Publishing}
}

@article{eraerds2010quantum,
  title={Quantum key distribution and 1 Gbps data encryption over a single fibre},
  author={Eraerds, Patrick and Walenta, Nino and Legr{\'e}, Matthieu and Gisin, Nicolas and Zbinden, Hugo},
  journal={New J. Phys},
  volume={12},
  number={6},
  pages={063027},
  year={2010},
  publisher={IOP Publishing}
}

@article{patel2014quantum,
  title={Quantum key distribution for 10 Gb/s dense wavelength division multiplexing networks},
  author={Patel, KA and Dynes, JF and Lucamarini, M and Choi, I and Sharpe, AW and Yuan, ZL and Penty, RV and Shields, AJ},
  journal={Appl. Phys. Lett.},
  volume={104},
  number={5},
  year={2014},
  publisher={AIP Publishing}
}

@article{khaksar2023simultaneous,
  title={Simultaneous BPSK classical communication and continuous variable quantum key distribution with a locally local oscillator regenerated by optical injection phase locked loop},
  author={Khaksar, Zeinab Sadat and Bahrampour, Alireza},
  journal={J. Laser Appl.},
  volume={35},
  number={4},
  year={2023},
  publisher={AIP Publishing}
}

@article{kumar2019experimental,
  title={Experimental demonstration of single-shot quantum and classical signal transmission on single wavelength optical pulse},
  author={Kumar, Rupesh and Wonfor, Adrian and Penty, Richard and Spiller, Tim and White, Ian},
  journal={Sci. Rep.},
  volume={9},
  number={1},
  pages={11190},
  year={2019},
  publisher={Nature Publishing Group UK London}
}

\bibliographystyle{naturemag}

%
\end{document}